\documentclass[aps,prl,reprint,floatfix]{revtex4-1}
\usepackage{blindtext}
\usepackage{amsmath}
\usepackage[dvipdfmx]{graphicx}
\usepackage[dvipdfmx]{color}
\usepackage{braket}
\usepackage{hyperref}
\usepackage{comment}
\usepackage{bm}
\usepackage{mathtools}
\usepackage{xcolor}

\definecolor{darkblue}{rgb}{0, 0, 0.5}

\hypersetup{
    colorlinks=true,
    linkcolor=darkblue,
    urlcolor=darkblue,
    citecolor=darkblue
}

\begin{document}
\title{Quantum Reservoir Computing Using Bose-Einstein Condensate with Damping}

\author{Yuki Kurokawa \textsuperscript{1}}
\author{Junichi Takahashi \textsuperscript{2}}
\author{Yoshiya Yamanaka \textsuperscript{3}}

\affiliation{
\mbox{${}^1$Graduate School of Information Science and Technology,
The University of Tokyo, 113-8656 Tokyo, Japan}\\
\mbox{${}^2$Faculty of Economics, Asia University, Tokyo 180-0022, Japan}\\
\mbox{${}^3$Institute of Condensed-Matter Science, Waseda University, Tokyo 169-8555, Japan}
}

\begin{abstract}
Quantum reservoir computing is a type of machine learning in which the high-dimensional Hilbert space of quantum systems contributes to performance. In this study, we employ the Bose-Einstein condensate of dilute atomic gas as a reservoir to examine the effect of reduction in the number of condensed particles, damping, and the nonlinearity of the dynamics. It is observed that for the condensate to function as a reservoir, the physical system requires damping.  The nonlinearity of the dynamics improves the  performance of the reservoir, while the reduction in the number of condensed particles 
degrades the performance.  
\end{abstract}

\maketitle

\section{Introduction}

Neural networks are mainly categorized into two types: feedforward neural networks (FNN) and recurrent neural networks (RNN).  
FNNs are structured with nodes arranged sequentially, allowing data to flow in one direction from the input to the output layer.  However, RNNs include a network architecture with feedback connections, which allow them to retain past input data as an internal memory, and are well-suited for learning and predicting sequential data.
However, RNNs are computationally expensive.
 To reduce the high training cost, the echo state network (ESN) was introduced, in which
only the weights of the network connected to the output nodes are updated, while the other weights are fixed \cite{ESN}.  The network part of an ESN serves as a memory and nonlinear transformation, and learning is performed by linear regression (LR) at the output nodes.  

Reservoir computing (RC) is a generalized computational framework that includes ESNs. In such frameworks, the input sequences are encoded into a high-dimensional space known as the reservoir, which stores past inputs and captures the characteristics of the data. 
Reservoirs are not confined to neural network models but are utilized in physical reservoir computing, which employs physical devices and systems \cite{water,octopus,octopus2,octopus3,electronic,spinwave,electronic_device,light,MZM}.
By using physical systems as reservoirs, the dynamics of physical systems can be substituted for calculations by computers and computational costs can be reduced.  In terms of the functionality, the nonlinear transformation within the physical reservoirs is facilitated by the nonlinear dynamics of the physical system,
while in network models, such a transformation is achieved through activation functions.
Similarly, memory formation in physical reservoirs is accomplished by hysteresis, whereas in network models, it is achieved by recurrent loops.

Quantum reservoir computing was proposed in Ref.\cite{isingreservoir}, in which the transverse Ising model was adopted as the reservoir.  
Although the dynamics of the model are linear, the reservoir functions due to the nonlinearity of expectation values of observables. Furthermore, it was demonstrated that the intrinsic degrees of freedom inherent to quantum systems
enable even a small number of qubits, specifically 5 to 7, to achieve performance comparable to that of network models comprised of 100 to 500 nodes. 
Recent research \cite{dampingreservoir} indicates that the noise causing amplitude damping improves the performance of quantum reservoirs. 
In the analysis in Ref.\cite{dampingreservoir}, a noisy intermediate-scale quantum (NISQ) device was employed as a quantum reservoir and the noise was uniformly introduced.

In this study, we employ a Bose-Einstein condensate system of cold atoms as the reservoir, which is a macroscopic quantum system with superfluidity. Our objective is to investigate the effect of damping in the system, which effectively arises from the escape of condensed particles, on the performance of the reservoir. We evaluate the performance quantitatively through the NARMA task.

\section{Bose-Einstein Condensate}

Bose-Einstein Condensation (BEC) is a phase transition, in which bosons condense into the lowest energy state \cite{BEC}.
  At extremely low temperatures, a large fraction of bosons occupy the lowest quantum state, resulting in macroscopic quantum phenomena such as superfluidity. 
 
In dilute cold atomic gases, BEC was first realized experimentally in 1995 \cite{Rb,Li,Na}. 
These gases are cooled using techniques such as laser cooling \cite{laser} and evaporative cooling \cite{evaporative}. 
Several techniques are crucial for the study and manipulation of cold atomic systems.
One of these is to control the effective coupling of the interatomic interaction by applying an external magnetic field, utilizing Feshbach resonance \cite{Feshbach}.
Another technique is non-destructive observation \cite{nondestructive}, which involves repeated measurements on the same condensate. 

As a theoretical framework, the condensate of dilute cold atoms can be well-described by two-body contact interactions.  
The dynamics of the condensate can be approximately described by the time-dependent Gross-Pitaevskii equation (TDGPE) \cite{pethick,pitaevskii,GPE}, 
\begin{equation}
\label{GPE}
i\hbar \frac{\partial \psi(\mathbf{r}, t)}{\partial t} = \left( -\frac{\hbar^2}{2M} \nabla^2 + V(\mathbf{r},t) + U_0 |\psi(\mathbf{r}, t)|^2 \right) \psi(\mathbf{r}, t) .
\end{equation}
Here, $\psi(\mathbf{r}, t)$ is the order parameter of the condensate, $V(\mathbf{r},t)$ is the potential, $U_0$ is the strength of the two-body interaction, and $M$ is the mass of the boson. The number of condensed particles is represented by $N=\int |\psi|^2 d\bm{r}$.
The TDGPE differs from the Schr\"{o}dinger equation as it includes a nonlinear term that arises from two-body contact interactions. 

Considering the damping of the condensate arising from the escape of condensed particles, 
we modify the TDGPE as in  \cite{dampingGP}:
\begin{multline}
\label{damping}
(1-i\gamma) \left\{ -\frac{\hbar^2}{2M}\nabla^2 +V(\bm{r},t)+U_0 |\psi(\bm{r},t)|^2 -\mu(t) \right\} \psi(\bm{r},t) \\=i\hbar \frac{\partial \psi(\bm{r},t)}{\partial t} .
\end{multline}
The parameter $\gamma \, (\gamma \geq 0)$ represents the damping strength, which leads to energy loss over time.
For the MIT experiment \cite{dampingGPexperiment}, it is estimated that $\gamma = 0.03$.
The chemical potential $\mu(t)$ is specified as 
\begin{equation}
\mu_N(t)= \frac {\int \psi(\bm{r},t)^* H(\bm{r},t) \psi(\bm{r},t) d\bm{r}}{\int |\psi(\bm{r},t)|^2 d\bm{r}} \, ,
\end{equation}
where, $H(\bm{r},t)=-\frac{\hbar^2}{2M}\nabla^2 +V(\bm{r},t)+U_0 |\psi(\bm{r},t)|^2$,
to ensure the conservation of the number of condensed particles.

\section{NARMA task}
The NARMA (Nonlinear AutoRegressive Moving Average) task \cite{NARMA} is used as a benchmark for the performance evaluation of RNNs or RCs to examine memory and nonlinearity.
This task is intended to emulate the following recurrence equation known as the NARMA system:
\begin{multline}
\label{NARMA}
d(n+1)=a_1 d(n)+a_2 d(n) \sum^{m-1}_{i=0} d(n-i)\\+a_3 u(n-m+1)u(n)+a_4 ,
\end{multline}
where, $d(n+1)$ is the output at step $n+1$, determined by the input sequence $ u(n-m+1)$ to $u(n) $ and the past output sequence $d(n-m+1)$ to $d(n)$.
The parameter $m$ indicates the order of the NARMA system, indicating the required memory steps.
This task requires the nonlinearity derived from the product terms $d(n) d(n-i)$ and $u(n-m+1)u(n)$.
Following the parameter settings described in \cite{isingreservoir}, we set $a_1=0.3, a_2=0.05, a_3=1.5, a_4=0.1$, and $m=10$, with the input $u(n)$ as a random number ranging from 0 to 0.2 with $5\times 10^3$ sequences.  
We divide the sequences into three parts. The first $10^3$ sequences are discarded to remove the transient state. The following $3 \times 10^3$ sequences are utilized as training datasets for supervised learning with linear regression (LR), using $d(n)$ directly calculated by Eq. (3). 
The last $10^3$ output sequences are employed for testing and evaluation.
Here, the performance is evaluated by the normalized mean square error (NMSE).

\section{Method}
In this study, we evaluate the performance of quantum reservoir computing using a cold atomic Bose-Einstein condensate with damping for the NARMA task. 
To simplify the calculations, we consider the one-dimensional condensate trapped in the harmonic potential:
\begin{equation}
V_\text{trap}(x)=\frac{1}{2} M \omega^2 x^2 ,
\end{equation}
where, $\omega$ is the strength of the harmonic oscillator. 
Hereafter, we set the units of length and time as $\sqrt{\hbar/M\omega}$ and $2/\omega$, respectively.

The procedure for evaluating the performance is as follows: First, the NARMA task is generated. 
Second, it is encoded into the condensate through the local Gaussian potential with the amplitude proportional to the input sequence of the NARMA task $u(n)$ at the center of the harmonic trap potential
\begin{equation}
V_{\mathrm{encode}}(x, t; n)=\alpha u(n) \exp^{-\frac{x^2}{2 \beta}},
\end{equation}
where, $\alpha$ and $\beta$ are the strength and width of the encoding potential, respectively.
Next, the TDGPE is solved using the second-order split-step Fourier method \cite{SSFM} and the density profile of the condensate $|\psi|^2$ is obtained for the output. 
Here,  we set the initial function of $\psi$ as the ground state when $V_{\mathrm{encode}}=0$.
 We divide the whole time for calculation into $5\times 10^5$ intervals and set its segment $\Delta t = 10^{-3}$. As there are $5\times 10^3$ steps of the NARMA task, each step is $100\times \Delta t$ long.

To differentiate the encoding of the $n$-th step from that of the $n+1$-th step easily, only the first $\Delta t$ of the time in one NARMA step is used for encoding, while during the remaining time, there is an absence of the encoding potential.
The domain for numerical calculation is $- 2^3 \leq x \leq 2^3$, and is divided into $2^{12}$ segments. 
As most of the condensate is in the area $-1 \leq x \leq 1$ and the system is symmetric about $x=0$, the density of the condensate $|\psi|^2$ is meaured in the area $0 \leq x \leq 1$.
To apply time multiplication \cite{isingreservoir}, in the $n$-th NARMA step, we measure $|\psi|^2$, 10 times per encoding at $t= 100n + 10k \Delta t ,\, k=0,\cdots, 9$, and obtain ten vectors $|\psi|^2_{n0}, \cdots, |\psi|^2_{n9}$.  At each step of the NARMA task, these ten vectors of different times are connected to one vector $\Phi_n \coloneqq \{|\psi|^2_{n0}, \cdots, |\psi|^2_{n9}\}$, which is the reservoir output of the $n$-th step of the NARMA task. 
 Finally, linear regression is performed on the acquired reservoir output, and $d(n)$ is predicted from $\Phi_n$.

\section{Performance Analysis}
\begin{figure}[ht]
\centering
\includegraphics[width=8.6cm]{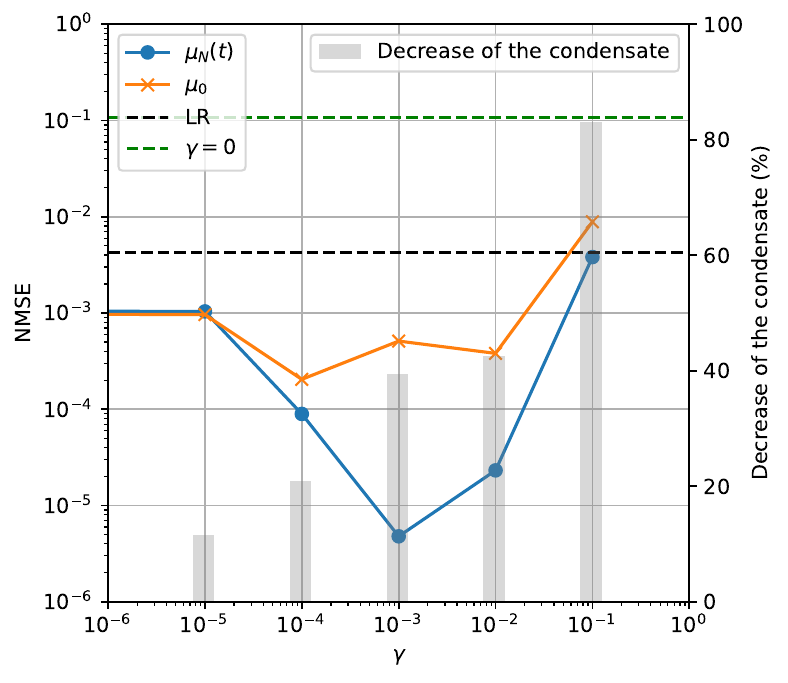}
\caption{\textbf{NMSE dependence on the damping parameter $\gamma$ at $U_0=0$.} The blue and orange solid lines show the NMSE of each $\gamma$ calculated using $\mu_N(t)$ and $\mu_0$, respectively. The green dashed line represents the NMSE at $\gamma=0$, as there is no difference of $|\psi|^2$ between $\mu_N(t)$ and $\mu_0$ at $\gamma=0$.  The black dashed line represents the NMSE obtained using only LR to expect $d(n+1)$ from $u(n)$ only without passing through the reservoir. The gray bar chart represents the percentage of decreased condensates when using $\mu_0$. }
\label{damping dependency}
\end{figure}

First, we evaluate the performance for each $\gamma$ parameter. 
To evaluate the effect of the decrease of the condensate, we examine two types of chemical potentials: 
$\mu_N(t)$, which is set to conserve the number of particles even with damping, and the constant chemical potential $\mu_0$ at the ground state when $V_\text{encode}=0$, which allows a decrease in the number of particles.
The results are shown in Fig.\ref{damping dependency}. Here, we fix the strength of the
two-body interaction $U_0 = 0$, which indicates the absence of the nonlinear term.

According to Fig.\ref{damping dependency}, 
the performance with the condensate reservoir without damping is actually inferior to that with only LR, which has just one step of memory and predicts $d(n+1)$ from $u(n)$ only.  
The moderate damping improves the performance, although excessive damping causes negative influence.  
In this experimental setting, the reservoir achieves the best performance at $\gamma=10^{-3}$. 
Furthermore, the performance is improved by setting the chemical potential to be time-dependent to keep the particle number constant. 
When we employ $\mu_0$, as $\gamma$ increases, the condensed particles decrease. At $\gamma=10^{-3}$, where the performance is the best, 40\% of the condensate decreases by employing $\mu_0$.
Therefore, we employ $\mu_N(t)$ in all of the following numerical experiments.

Next, we evaluate the performance dependence on the interaction between particles $U_0$.
\begin{figure}[ht]
\centering
\includegraphics[width=8.6cm]{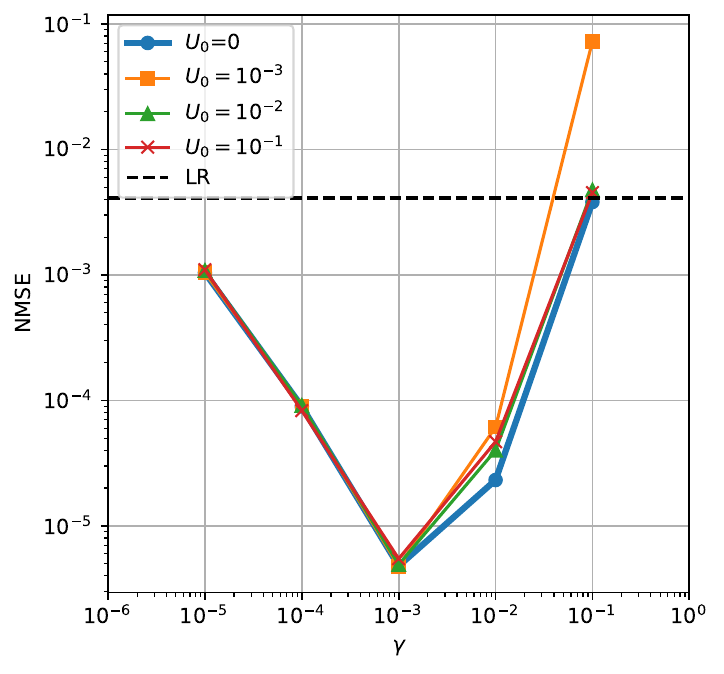}
\caption{\textbf{NMSE dependence on the interaction parameter $U_0$.}  The blue, orange, green, and red solid lines show the NMSE of each $\gamma$ at $U_0=0,10^{-3},10^{-2}$ and $10^{-1}$, respectively.  The black dashed line represents the NMSE obtained using only LR to expect $d(n+1)$ from $u(n)$ only without passing through the reservoir.  }
\label{nonlinearity dependency}
\end{figure}

As shown in Fig.\ref{nonlinearity dependency}, for any $U_0$, the performance peaks at $\gamma=10^{-3}$.
In the range $\gamma<10^{-3}$, the difference in performance caused by nonlinearity is negligible, while in the range $\gamma> 10^{-3}$, the nonlinearity causes negative influence. 
Overall, the impact of the damping is significantly stronger compared to the impact of nonlinearity,

Finally, we evaluate the performance dependency on $U_0$ in detail at $\gamma=10^{-3}$.
\begin{figure}[ht]
\centering
\includegraphics[width=8.6cm]{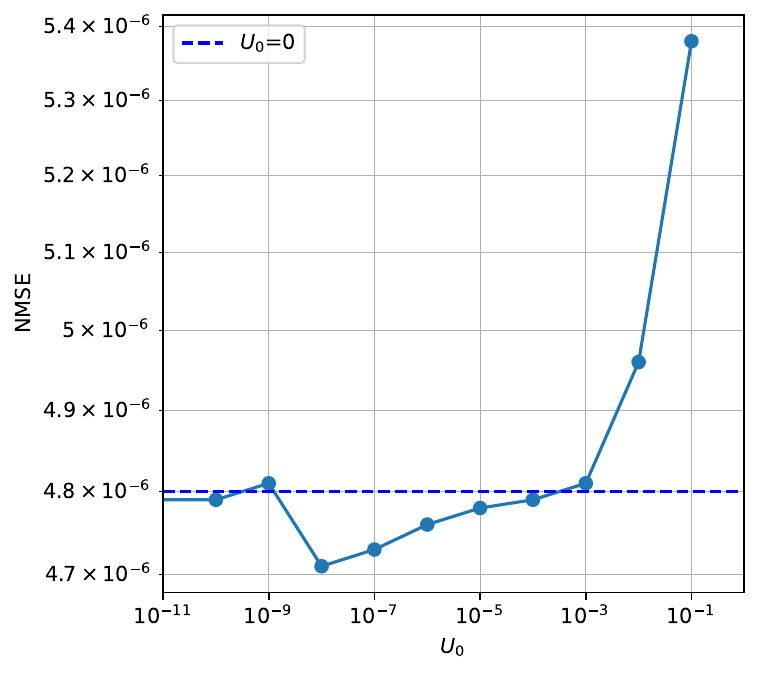}
\caption{\textbf{NMSE dependence on the interaction parameter $U_0$ at $\gamma=10^{-3}$.}  The blue dashed line is the NMSE of $U_0=0$. The blue solid line is the NMSE of different $U_0$.  }
\label{observe dependency}
\end{figure}\noindent
As illustrated in Fig.3, when $U_0<10^{-8}$, the nonlinearity $U_0$ makes little difference from $U_0=0$. The lowest NMSE is observed at $U_0=10^{-8}$. Conversely, when $U_0$ exceeds $10^{-8}$, increased nonlinearity detrimentally affects the performance.
 The positive effect is dominant for $U_0 \leq 10^{-8}$, while the negative effect is dominant for $U_0 > 10^{-8}$.

\section{discussion}
We address five key points regarding the results presented thus far. First, the decrease in the condensate causes performance degradation because it causes the reservoir to be in a transient state. This transient state creates a discrepancy between the reservoir's condition during the expectation phase and during the learning phase, leading to poor performance.

Second, damping is essential for the reservoir because without it, all inputs are stored indefinitely within the reservoir.
Such indefinite storage makes it difficult to extract the required inputs and their nonlinear transformations, which consist of the most recent $m$ steps. 
Therefore, damping helps to select these critical inputs, ensuring that the reservoir functions effectively.

Third, while damping is crucial, excessive damping also leads to performance degradation because it removes some of the most recent $m$ step inputs, which are vital for the outputs.

Fourth, the impact of nonlinearity on the performance is less significant than that of damping. This is due to the fact that the nonlinear transformation is already accomplished through the calculation of $|\psi|^2$. 
This result agrees with the previous research \cite{isingreservoir}, according to which the nonlinearity of the expectation values allows a quantum system to function effectively despite its linear dynamics. Our research quantitatively determines whether the nonlinearity derived from observation is sufficient for the system's requirements.

Finally, excessive nonlinear dynamics have a negative effect because the nonlinearity stems from the repulsive interaction between particles. As the nonlinearity increases, the condensate becomes more diffused, causing parts of it to move out of the observation area, which in turn negatively impacts the performance. Thus, although both damping and nonlinearity are crucial, their optimal control is necessary to achieve the best performance.

\section{summary}
In this study, it is observed that appropriate damping and nonlinearity of the dynamics of the physical system improve the performance of the reservoir, while the decrease in the number of condensed particles degrades the performance.
Although performance improvement due to damping has been noted in previous research using the noisy intermediate-scale quantum (NISQ) device \cite{dampingreservoir}, our numerical results using the Bose-Einstein condensate as the reservoir, also support the some observation regarding performance improvement.
This study elucidates the conditions under which physical systems can function effectively as reservoirs.
In the future, we will evaluate the performance of quantum reservoir computing using condensates and analyze the effects of quantum phenomena such as interference, tunneling, and superposition.

\end{document}